\documentclass[
aps,
prl,
superscriptaddress,
reprint,
nofootinbib 
]{revtex4-1}

\usepackage[utf8]{inputenc} 
\usepackage{hyperref} 
\usepackage{amsmath} 
\usepackage{amsfonts}
\usepackage{stfloats}
\usepackage{amsthm}
\usepackage{amssymb}
\usepackage{graphicx,nicefrac} 
\usepackage{lmodern} 
\usepackage{braket} 
\usepackage{subcaption} 
\usepackage{color} 
\usepackage[normalem]{ulem} 

\graphicspath{{./}{./Plots/}}

\newcommand{\te}{\textemdash}
\newcommand{\tr}{\text{Tr}}

\newcommand{\abs}[1]{\left|#1\right|}
\newcommand{\Seff}{S_\text{eff}}
\newcommand{\SLO}{S_\text{LO}}
\newcommand{\SNLO}{S_\text{NLO}}
\newcommand{\SNNLO}{S_\text{NNLO}}
\newcommand{\td}{_\text{3d}}
\newcommand{\pmin}{\phi_\text{min}}
\newcommand{\yLO}{y_\text{LO}}

\newcommand{\sectioninline}[1]{\textit{#1}.---} 

\newcommand{\Uppsala}{\affiliation{
		Department of Physics and Astronomy, Uppsala University,
		Box 516, SE-751 20 Uppsala,
		Sweden
}}

\newcommand{\Hamburg}{\affiliation{
		II. Institute of Theoretical Physics, Universität Hamburg, D-22761, Hamburg, Germany
}}

\newcommand{\DESY}{\affiliation{
		Deutsches Elektronen-Synchrotron DESY, Notkestr. 85, 22607 Hamburg, Germany
}}

\begin{document}
	
	\title{Convergence of the nucleation rate for first-order phase transitions}
	\date{\today}
	
	\author{Andreas Ekstedt}
	\email{andreas.ekstedt@desy.de}
	\Uppsala \Hamburg \DESY

	\begin{abstract}
		This paper investigates the importance of radiative corrections for first-order phase transitions, with particular focus on the bubble-nucleation rate. All calculations are done with a strict power-counting, and observables are consistently calculated at every order. This ensures that physical quantities are gauge and renormalization-scale invariant. Furthermore, to avoid large logarithms at high-temperatures, an effective three-dimensional theory is used. This effective theory automatically incorporates higher-order thermal masses. The results of this paper indicate that sub-leading corrections to the rate can be large. This is partly because radiative corrections are enhanced for large bubbles. To illustrate the calculations, three models are considered: a real-scalar model, a radiative-barrier model, and a model with an effective dimension $6$ operator. Relevant observables are calculated for each model, and the reliability of perturbation theory is discussed.
	\end{abstract}
	
	\maketitle

 \fontdimen2\font=0.6ex
\spaceskip=0.3em
\thickmuskip=5mu
\medmuskip=4mu minus 2mu
\setlength{\parskip}{0pt}

\sectioninline{Introduction}
A cosmological first-order phase transition is a watershed moment. If such a transition occurred, not only would it leave a trail of gravitational waves~\cite{Hindmarsh:2017gnf,Vaskonen:2016yiu,Grojean:2006bp,Giese:2020znk}, but it could also explain the observed Baryon asymmetry through Electroweak Baryogensis~\cite{Kuzmin:1985mm}. Therefore, with the advent of gravitational-wave cosmology, the community\te theoretical and experimental\te prepare for upcoming experiments~\cite{NANOGrav:2020bcs, LISA:2017pwj, DECIGO:2011zz, Taiji:2018tsw, AEDGE:2019nxb}. The success of which promises an unique window into the early universe, and the chance to not only confirm a first-order transition, but to probe the Higgs potential itself~\cite{Friedrich:2022cak,Ramsey-Musolf:2019lsf}.

Yet for this to pass, theoretical tools must be up to par.
Unfortunately, there are signifiant theoretical uncertainties for non-equilibrium processes like bubble nucleation~\cite{Gould:2021oba,Caprini:2019egz,Guo:2021qcq}. These uncertainties in turn prevent reliable predictions of the gravitational-wave spectrum. Thus limiting any attempt to constrain the underlying physics if a signal is detected.

So it is crucial to improve computational methods; these are divided into numerical lattice computations~\cite{Moore:2000jw,Gould:2022lat} and perturbation theory~\cite{Hirvonen:2021zej,Lofgren:2021ogg,Gould:2021ccf,Croon:2020cgk,Gould:2021oba}.
Although lattice computations are preferable, they are slow even for equilibrium observables. This leaves perturbation theory as the only viable option for studying models with many free parameters. Accordingly, the accuracy of perturbative calculations must be known. 

To that end, this paper endeavours to give reliable predictions for bubble-nucleation; to calculate the nucleation rate beyond leading order; to determine when perturbation theory breaks down; and to pave the way for precise predictions of the gravitational-wave spectrum.

\sectioninline{High-temperature calculations}
Physical quantities are renormalization-scale invariant~\cite{Gould:2021oba}, gauge invariant~\cite{Buchmuller:1994vy,Hirvonen:2021zej}, and free from infrared divergences~\cite{Laine:1994bf, Laine:1994zq} in a consistent perturbative expansion.

However, a naive loop-expansion does not work at high temperatures. This is because loop corrections are enhanced, which for example leads to thermal mass corrections. In addition, calculations typically contain large logarithms if the relevant energy-scale is much smaller than the temperature.

These issues can be solved by integrating out high-energy fluctuations and working with an effective field theory (EFT). This theory describes energy-scales relevant to the phase transition, and all temperature dependence is contained in effective couplings~\cite{Croon:2020cgk,Kajantie:1995dw,Farakos:1994kx}.

Due to the universality of effective field theories, this makes it possible to study an entire class of zero-temperatures theories with a single EFT.

\sectioninline{The nucleation rate}
It is possible to use classical nucleation theory to calculate the rate of nucleating bubbles at high temperatures~\cite{Langer:1969bc,Linde:1981zj}.
Physically the nucleation-rate is controlled by a Boltzmann factor, which captures the probability for a thermal transition between two phases. Other dissipative effects, such as damping~\cite{Langer:1969bc}, are omitted in this paper as they are suppressed.

Consider first a particle trapped in a potential well, where the probability to escape is controlled by the height of the barrier separating the two minima: $\Gamma \sim e^{-V/T}$. In field-theory the barrier-height is replaced by an action. This action is evaluated on a classical solution to the equations of motion, the bounce~\cite{Coleman:1977py,Linde:1981zj}.

To illustrate the form of the rate, take a scalar field: the nucleation rate is proportional to
\begin{align}\label{eq:LORate}
	\Gamma\propto e^{-S_B}.
\end{align}
If we assume that the scalar potential $V(\phi)$ has one minima at $\phi=\phi_\text{FV}$, and a deeper minima at $\phi=\phi_\text{TV}$, the bounce is given by
\begin{align}\label{eq:LOBounce}
	&\nabla^2\phi_B(\abs{\vec{x}})=V'[\phi_B(\abs{\vec{x}})],\quad \lim_{\abs{\vec{x}}\rightarrow \infty}\phi_B(\abs{{\vec{x}}})=\phi_\text{FV},\nonumber
	\\& \left.\vec{\nabla}\phi_B(\abs{\vec{x}})\right|_{\abs{\vec{x}}=0}=0,
\end{align}
and the bounce action is
\begin{align}\label{eq:LOBounceAction}
	S_B=\int d^3x \left[\frac{1}{2} (\vec{\nabla}\phi_B)^2+V[\phi_B]\right].
\end{align}
Note that $S_B$ is three dimensional. Furthermore, as mentioned, we here consider an effective high-temperature theory. This means that all masses and couplings implicitly depend on the temperature, and that we work with an effective three-dimensional theory~\cite{Gould:2021ccf}.

Higher-order corrections to the rate come from including fluctuations around the bounce solution.
For example, the next-to-leading order (NLO) result is~\cite{Linde:1981zj,Langer:1969bc}
\begin{align}
	\Gamma \propto \prod_i \det\left[-\nabla^2+M^2_i[\phi_B]\right]^{-1/2} e^{-S_B}.
\end{align}
This functional determinant depends on the leading-order bounce through field-dependent masses \footnote{The rate should be normalized by a corresponding determinant evaluated at $\phi=\phi_\text{FV}$.}.  

Though the determinant is formally sub-leading, it can be comparable to the exponent. To see when this happens, it is useful to rewrite the rate as
\begin{align}
	\Gamma \propto e^{-\Seff[\phi_B]},\quad \Seff=S_B+\SNLO,
\end{align}
where we have defined the one-loop effective action~\cite{Weinberg:1992ds}
\begin{align}\label{eq:SNLODet}
	\SNLO[\phi_B]=\frac{1}{2} \sum_i \tr \log  \left[-\nabla^2+M^2_i[\phi_B]\right].
\end{align}

If we consider large bubbles with radius $R$, the leading-order bounce action scales as $S_B\sim R^2$~\cite{Coleman:1977py}, while the NLO action scales as
\begin{align}\label{eq:ThinWallBehav}
	\SNLO\sim -R^3 \sum_i \left[(M^2_i[\phi_\text{TV}])^{3/2}-(M_i^2[\phi_\text{FV}])^{3/2}\right].
\end{align}
We see that for large enough $R$, $\SNLO$ overpowers $S_B$, and perturbation theory breaks down. In addition, equation \ref{eq:ThinWallBehav} indicates that higher-order corrections are enhanced even for medium-sized bubbles.

To systematically study higher-order corrections, it is necessary to introduce a powercounting. As an example, consider the potential
\begin{align}\label{eq:RadPotential}
	V(\phi)=\frac{1}{2}m^2\td \phi^2-\frac{1}{16\pi}g^3\td  \phi^3+\frac{1}{4}\lambda\td\phi^4.
\end{align}
This potential appears in the Standard model when the physical Higgs mass is small $m_H\sim 40~\text{GeV}$. In that case $\phi=\left\langle \Phi \right\rangle$ where $\Phi$ is an $\mathrm{SU}(2)$ doublet, and the cubic term in equation \ref{eq:RadPotential} arises from vector-boson loops~\cite{Arnold:1992rz}. As such, the potential \ref{eq:RadPotential} is said to describe a radiative barrier. It should be noted that the same potential also describes a real-scalar theory, which can be seen by redefining the gauge coupling: $g_{3d}^3\rightarrow \eta\td$~\cite{Gould:2021dzl,Baacke:1993ne}. Where $\eta\td$ is a cubic coupling-constant.

At leading order, the effective couplings in equation \ref{eq:RadPotential} depend on the original zero-temperature couplings schematically as~\cite{Kajantie:1995dw}
\begin{align}
	\lambda\td =T \lambda, \quad g\td^2=T g^2,\quad m\td^2=m^2+a T^2,
\end{align}
where $a$ is a function of zero-temperature couplings.

We first consider the radiative-barrier case in an $\mathrm{SU}(2)$ model. The cubic term in equation \ref{eq:RadPotential} comes from integrating out vector bosons. For this to be consistent, the vector-boson mass must be parametrically larger than the Higgs mass: $\frac{m_H^2}{m_A^2}\sim \frac{\lambda\td}{g\td^2}\sim \frac{\lambda}{g^2}\ll 1$. This encourages us to define the dimensionless couplings
\begin{align}\label{eq:defOfx}
	x\equiv \frac{\lambda\td}{g\td^2}, \quad y\equiv \frac{m\td^2}{g\td^4}.
\end{align}
Where in addition to $x$, we also use the dimensionless variable $y$~\cite{Kajantie:1995kf}.
With these variables, and a field/coordinate rescaling, the potential is
\begin{align}\label{eq:LOPotential}
	V(\phi)=\frac{1}{2}y \phi^2-\frac{ 1 }{16\pi}\phi^3+\frac{1}{4}x \phi^4.
\end{align}

Different minima of $V(\phi)$ correspond to different phases. In equation \ref{eq:LOPotential} one minimum is at $\phi=0$, and another at $\phi=\pmin \neq 0$. We say that a phase transition can first occur when $\Delta V\equiv V(\pmin)-V(0)=0$. And because the problem only depends on two variables, it is useful to define the critical mass as $\Delta V(y_c,x)=0$~\cite{Croon:2020cgk,Ekstedt:2022hep}. For a specific zero-temperature model it is then possible to find the critical temperature given $y_c$. But we refrain from doing so to keep the setup general.

Analogously, one can define the nucleation mass $y_N$ as the solution of $\Seff(y_N,x)=126$. This is the condition for roughly two-thirds of the Universe to be in the broken phase~\cite{Guo:2021qcq,Caprini:2019egz}.

Close to the nucleation mass, the perturbative expansion is organized as
\begin{align}\label{eq:EffectiveAction}
	\Seff= \SLO+ x \SNLO+x^{3/2}\SNNLO+\ldots
\end{align}
where powers of $x$ only denote how the terms scale. In this expansion, integer powers of $x$ come from vector loops, and rational powers of $x$ come from scalar loops. It should be noted that the expansion in equation \ref{eq:EffectiveAction} changes for small $x$; as we shall see, for $x \lesssim10^{-2}$ the $\SNLO$ contribution is suppressed by a factor $\sim6.6$ relative to $\SLO$ irrespective of $x$.

In equation \ref{eq:EffectiveAction}, the leading-order action $ \SLO$ is defined by equations \ref{eq:LOBounce}, \ref{eq:LOBounceAction}, and \ref{eq:LOPotential}; while $\SNLO$ comes from integrating out vector bosons~\cite{Ekstedt:2021kyx,Lofgren:2021ogg,Hirvonen:2021zej}:
\begin{align}\label{eq:NLOAction}
	\SNLO=\int d^3x  &\left\lbrace
	-\frac{11}{32\pi }\frac{(\partial_\mu \phi_B)^2}{\phi_B}
	\right.
	\\& +\left. 
	\frac{\phi_B ^2}{ (4\pi) ^2}  \left(-\frac{51}{32} \log \frac{\phi_B}{\mu_3} -\frac{63}{32} \log \frac{3}{2} + \frac{33}{64}\right) \right\rbrace,\nonumber
\end{align}
where $\phi_B$ is the leading-order bounce solution, and we take $\mu_3=1$. One finds that $\SNLO$ scales as $y^{1/2}$, so it is suppressed by a power of $y$ relative to $\SLO$.

Finally, $\SNNLO$ is given by 
\begin{align}\label{eq:NNLOAction}
\frac{1}{2} \left\lbrace \tr \log  \left[-\nabla^2+M_H^2\right]+3  \tr \log  \left[-\nabla^2+M_G^2\right] \right\rbrace.
\end{align} The first term comes from Higgs bosons, and the second from Goldstone bosons. Their field-dependent masses are given by~\cite{Ekstedt:2022hep,Ekstedt:2020abj}
\begin{align}
	M_H^2=V''[\phi_B],\quad M_G^2=\phi_B^{-1}V'[\phi_B].
\end{align}

In the real-scalar model we have neither Goldstone nor vector bosons. Thus we should set $\SNLO=0$ and omit the Goldstone contribution in equation \ref{eq:NNLOAction}.

The leading-order action has to be found numerically, yet it can be approximated by the expression~\cite{Dine:1992wr,Ekstedt:2021kyx}
\begin{align}\label{eq:LOActionFit}
	S_\text{LO}=\kappa \left[7.24+5.68 \gamma +\frac{10.4}{1-\gamma }+\frac{1.25}{(1-\gamma )^2} \right],
\end{align}
where $\kappa=64\pi^2y^{3/2}$ and $ \gamma =128\pi^2 x y$. 

For the radiative barrier, direct calculations show that both  $\SNLO$ and $\SNNLO$ scale as $(1-\gamma)^{-3}$ when $\gamma \rightarrow 1$ \footnote{The bubble radius is $R\propto (1-\gamma)^{-1}$.}. This corresponds to the thin-wall limit: $y\rightarrow y_c=\frac{1}{128\pi^2 x}$.
In addition, for smaller $\gamma$ both $\SNLO$ and $\SNNLO$ are of similar size as $S_\text{LO}$, without the factor of $\kappa$~\cite{Ekstedt:2021kyx}.

There are then two cases when the expansion breaks down:  $\kappa \rightarrow 1$ and $\gamma \rightarrow 1$.

To estimate when $\kappa=1$, note that the nucleation mass is always lower than the critical mass, and because $\kappa \propto y^{3/2}$, we want $y$ to be as large as possible. Putting these observations together, we expect that the expansion breaks down (for sure) when 
\begin{align}\label{eq:xLowerBound}
	64\pi^2y_c^{3/2}=1 \implies x=\frac{1}{8 \pi^{2/3}}\approx 0.058
\end{align}
This is an upper bound on $x$ that applies to both the radiative-barrier and the real-scalar model; similar bounds appear in other models.

For the real-scalar model one finds that $\SNNLO$ grows as $(1-\gamma)^{-2}$, so there are no $\gamma \rightarrow 1$ problems. In contrast, for the radiative barrier both $\SNLO$ and $\SNNLO$ grow as $(1-\gamma)^{-3}$, which means that radiative corrections are larger.

\sectioninline{Observables}
The nucleation mass is defined by\footnote{There are different definitions of the nucleation temperature in the literature~\cite{Croon:2020cgk,Caprini:2019egz}. However, using a different definition of the nucleation mass does not qualitatively change the results.}
\begin{align}\label{eq:NucleationMass}
	\left[S_\text{LO}+x S_\text{NLO}+\ldots\right]_{y=y_N}=126.
\end{align}
This equation can be solved by expanding $y_N$ in powers of $x$:
\vspace*{-0.3cm}
\begin{align}
	y_N=y_\text{LO}+x y_\text{NLO}+x^{3/2} y_\text{NNLO}\ldots
\end{align}
The solution to NLO is
\begin{align}
	& S_\text{LO}\vert_{y=y_\text{LO}}=126, \quad  y_\text{NLO}=-\left.\frac{ S_\text{NLO }}{\partial_y S_\text{LO }}\right|_{y=y_\text{LO}}.
\end{align}
In general $\yLO$ needs to be found numerically, but we can still study some limits analytically. Indeed, from equation \ref{eq:LOActionFit} we see that  $\yLO$ grows as $x^{-1}$ for large $x$, and $\yLO \approx  0.048 $ for small $x$. One then finds that $\kappa$ can not be larger than $\kappa \approx 6.6$, which means that it is not possible to make higher-order corrections arbitrarily suppressed.

Given $y_N$, we can calculate observables such as the (inverse) phase-transition duration, which is given by~\cite{Croon:2020cgk,Caprini:2019egz}
\begin{align}
	\beta_N/H_N=\left. \frac{d}{d\log T}\Seff\right\vert_{y=y_N}.
\end{align}
Because our couplings implicitly depend on the temperature, we can use the chain rule to express $\beta_N$ in terms of $x$ and $y$ derivatives of $\Seff$~\cite{Gould:2019qek}. In addition, since
\begin{align}
	\frac{d }{d \log T} y \gg  \frac{d }{d \log T}x,
\end{align}
we can approximate~\cite{Gould:2019qek}
\begin{align}
	\beta_N/H_N \approx \left.\frac{d y}{d \log T}\nabla_y \Seff \right|_{y=y_N}.
\end{align}
Furthermore, expanding everything in powers of $x$ we find to NLO
\begin{align}
	\left.\nabla_y \Seff \right|_{y=y_N}= \nabla_y \SLO+x\left[y_\text{NLO}\nabla^2_y \SLO+\nabla_y \SNLO \right]\nonumber
\end{align}
where all terms are evaluated at $y=\yLO$. Note that $\nabla_y \Seff$ is calculable purely within the effective theory, while $\frac{d y}{d \log T}\sim 4$ depends on the original zero-temperature model.

The strict perturbative expansion is not only simple, but also renormalization-scale and gauge invariant at every order~\cite{Ekstedt:2022hep,Hirvonen:2021zej}.

\sectioninline{A dimension $6$ operator}
Consider now a model with a leading-order potential
\begin{align}\label{eq:TreePot}
	V(\phi)=\frac{1}{2}m^2\td \phi^2-\frac{1}{4}\lambda\td \phi^4+\frac{1}{32} c_6 \phi^6.
\end{align}
The normalization is chosen so that the broken and unbroken minima coincide when $c_6=\frac{\lambda\td^2}{m^2\td}$.

This model appears, for example, when effective operators are added to the Standard Model~\cite{Croon:2020cgk}: we consider this to be the case here. That is, neglecting the hypercharge coupling, we consider an $\mathrm{SU}(2)$ gauge theory with a doublet scalar $\left\langle\Phi\right\rangle =\phi$. The three dimensional $c_6$ coupling is related to the zero-temperature one (at leading-order) via $c_6=T^2 c_{6,4d}$.

To study this model it is useful to introduce the dimensionless coupling
\vspace*{-0.2cm}
\begin{align}
	y=\frac{m^2\td}{\lambda\td^2}.
\end{align}
At leading-order everything depends on $y$ and $c_6$; both scalars and vector-bosons contribute at NLO according to equation \ref{eq:SNLODet}. These contributions must be calculated numerically~\cite{Ekstedt:2021kyx}, but we note that the vector-boson contribution grows as $x^{-3/2}$ for small $x$, where $x$ was defined in equation \ref{eq:defOfx}.

The tree-level action can be approximated by~\cite{Ekstedt:2021kyx}
\begin{align}\label{eq:LOActionDim6}
	\SLO=\sqrt{y}\left[1.76-0.142 \gamma+\frac{12.6}{(1-\gamma)} +\frac{4.19}{(1-\gamma)^2} \right],
\end{align}
where $\gamma =c_6 y$. 

As before, we can determine $y_N$ and $\beta_N$ in powers of $c_6$ and $x$. 

\begin{figure}[h!]
	\begin{subfigure}{0.5\textwidth}
		\includegraphics[width=1\textwidth]{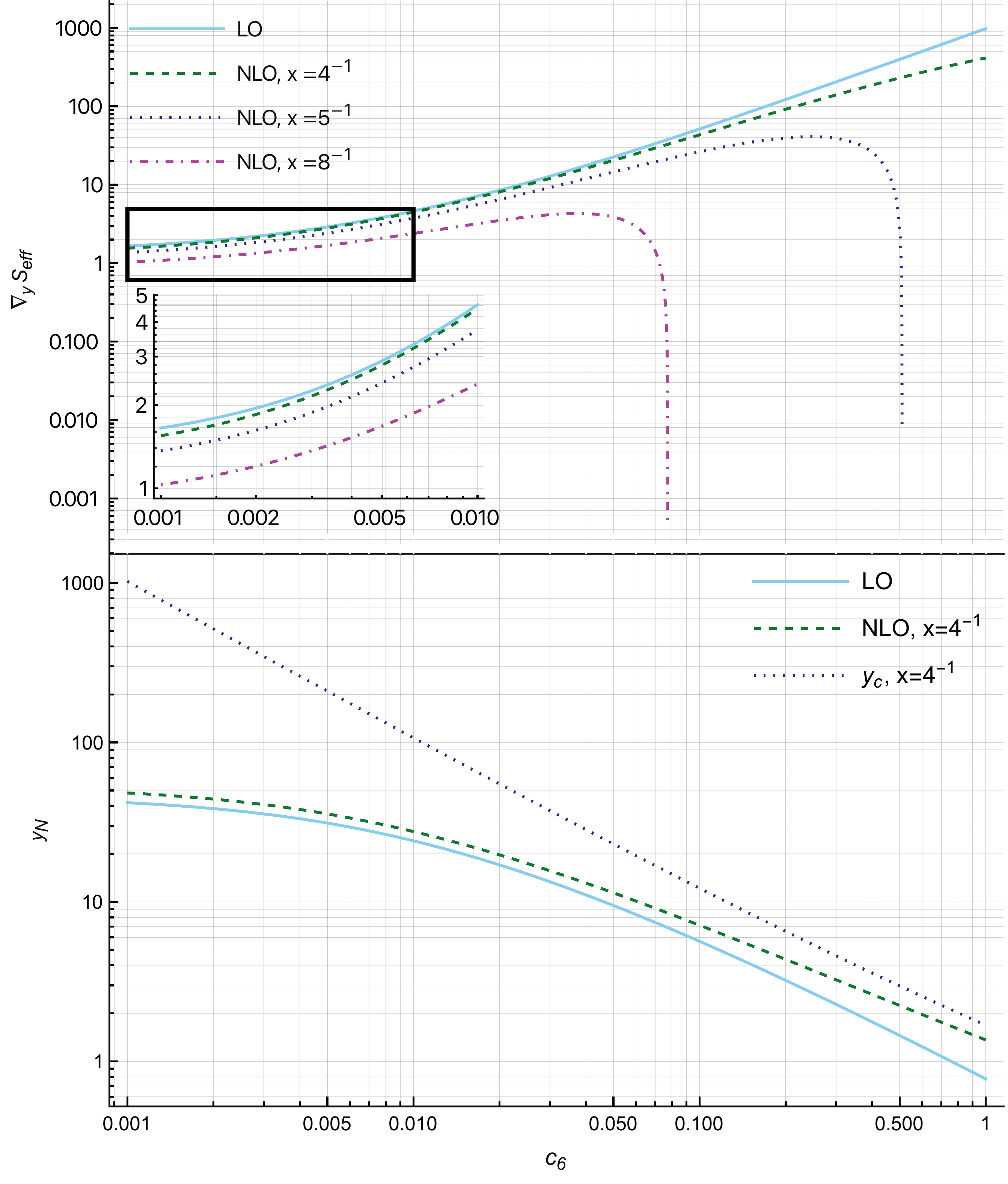}
	\end{subfigure}
	\caption{The lower plot shows the nucleation mass $y_N$ as a function of $c_6$ for the potential \ref{eq:TreePot} with $x=4^{-1}$. The critical mass (to NLO) is shown for comparison. The upper plot shows $\nabla_y \Seff\propto \beta_N$ at the nucleation mass.}
	\label{fig:yNTL}
\end{figure}

Following the same arguments as for the radiatively-induced potential, we expect that perturbation theory becomes unreliable when $c_6\gtrsim 1$. However, because the NLO action grows as $x^{-3/2}$ for small $x$, this bound is modified to $c_6\gtrsim x^{3}$. Furthermore, since NLO contributions are enhanced by large-bubble effects, the actual bound is slightly lower as shown in figure \ref{fig:yNTL} .

There is also an absolute lower-bound on $x$ regardless of the value of $c_6$. This is because $y_N$ can not be arbitrarily large. Indeed, using equation \ref{eq:LOActionDim6} we see that $y_N$ is largest when $\gamma=0$, which corresponds to $\yLO\approx 46$. And because the leading-order action scales as $\sqrt{y}$, it is not possible to consider arbitrarily small values of $x$. Numerically one finds that perturbation theory does not work for $x$ smaller than $x\sim 10^{-1}$.

\begin{figure}[h!]
	\begin{subfigure}{0.4\textwidth}
		\includegraphics[width=1.0\textwidth]{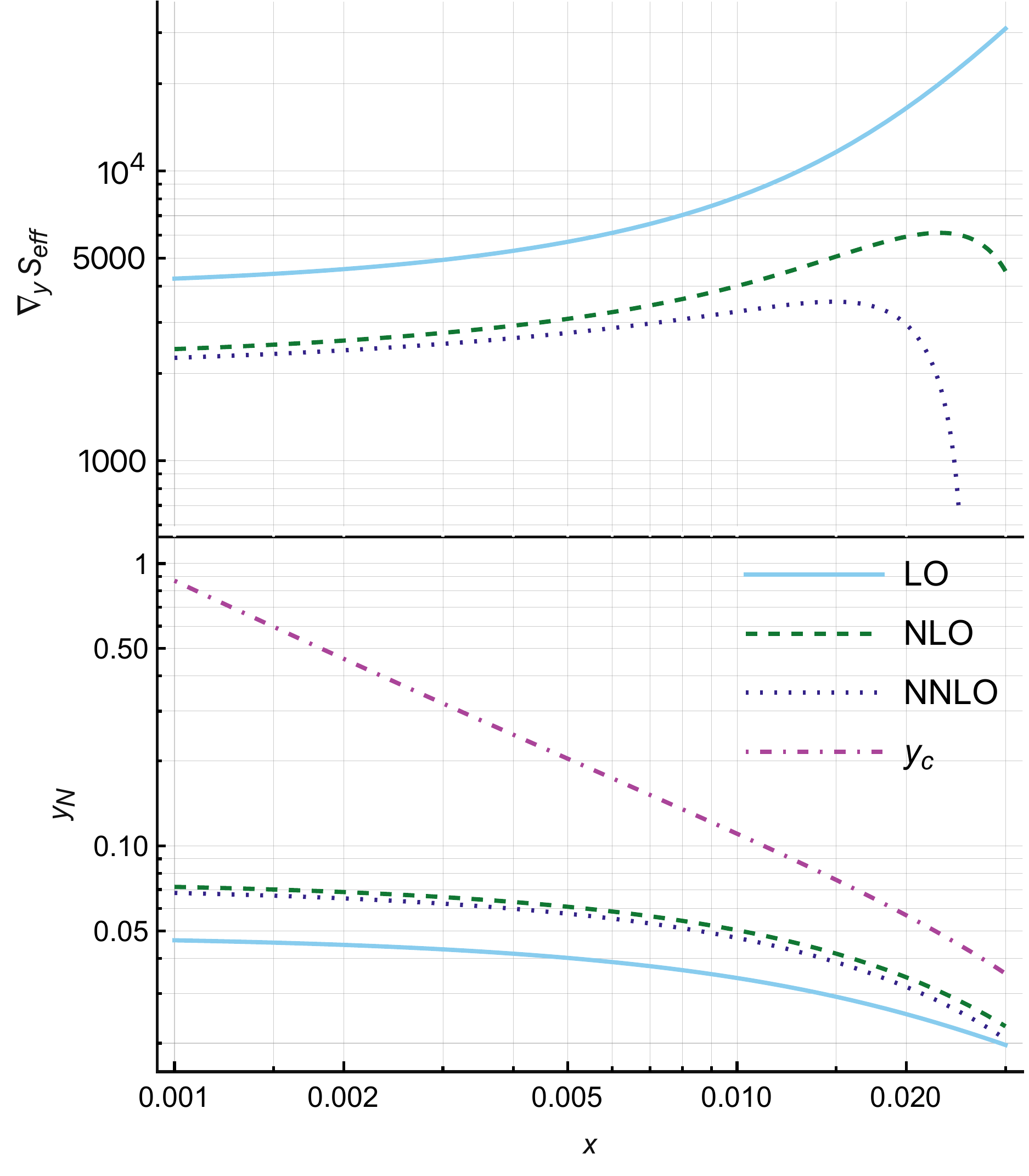}
	\end{subfigure}
	\caption{The lower plot shows the nucleation mass $y_N$ as a function of $x$. The critical mass (to NNLO) is shown for comparison. The upper plot shows $\nabla_y \Seff\propto \beta_N$ at the nucleation mass. }
	\label{fig:yN}
\end{figure}

\begin{figure}[h!]
	\begin{subfigure}{0.4\textwidth}
		\includegraphics[width=1.0\textwidth]{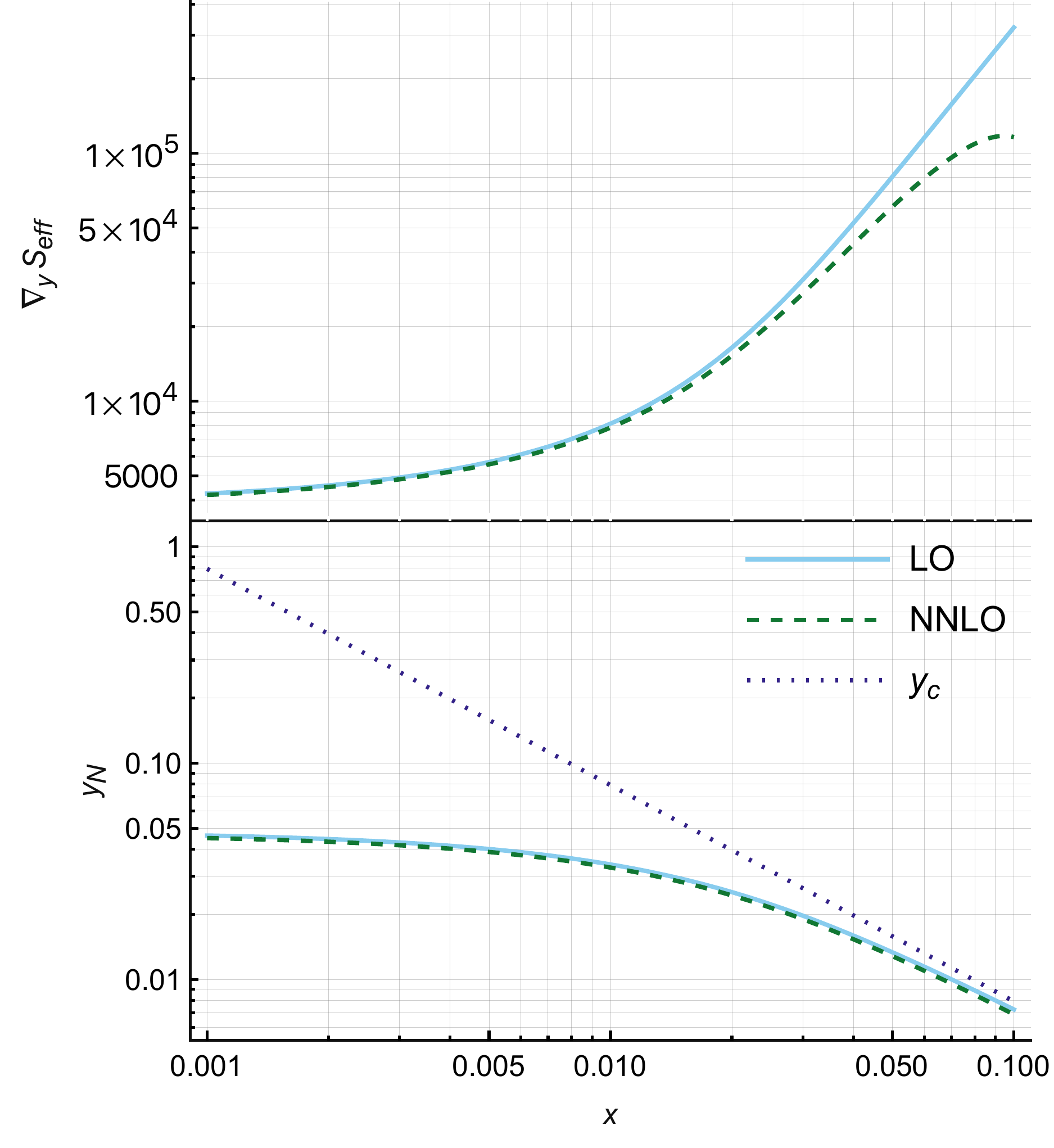}
	\end{subfigure}
	\caption{The lower plot shows the nucleation mass $y_N$ as a function of $x$ for the real-scalar model. The critical mass (to NNLO) is shown for comparison. The upper plot shows $\nabla_y \Seff\propto \beta_N$ at the nucleation mass. There is no NLO contribution for this model.}
	\label{fig:yNRS}
\end{figure}
\sectioninline{Results}
The radiatively-induced potential is defined in equation \ref{eq:RadPotential}, and the results are shown in figure \ref{fig:yN}. We see that as the nucleation mass decreases\te meaning a weaker transition\te perturbation theory breaks down. This is expected because weak transitions are generally non-perturbative~\cite{Moore:2000jw,Gurtler:1997hr,Kajantie:1995kf}, yet this breakdown occurs already at $x\approx 0.02$, instead of the bound derived in equation \ref{eq:xLowerBound}. This is because $\SNLO$ is numerically large, and since both $\SNLO$ and $\SNNLO$ are enhanced for large bubbles. Note that radiative corrections are large even for smaller $x$ where the expansion is expected to perform well. Indeed, the NLO result for $\nabla_y \Seff$, and thus $\beta_N$, is roughly a factor of two smaller than the leading-order result. Still, there is not another large jump once $\SNNLO$ is included. This indicates that  higher-order corrections can be large without invalidating the perturbative expansion.

Radiative corrections are smaller for the real-scalar case as shown in figure \ref{fig:yNRS}. In this model the expansion gets worse around $x\approx 0.05$ as expected from equation \ref{eq:xLowerBound}. The modest size of the NNLO correction stems from that the Higgs mass is equal in the broken and true minimum; which coupled with equation \ref{eq:ThinWallBehav}, shows that there is no (large-bubble) $R^3$ enhancement.

The results for a Standard-Model like potential with an effective $\phi^6$ operator are shown in figure \ref{fig:yNTL}. The results indicate that the expansion breaks down for small $x$. This is because  $\SNLO$ grows as $x^{-3/2}\sim \lambda\td^{-3/2}$, which pushes the range of validity to small $c_6$ values: $c_6 \lesssim x^3$.

Interestingly, even when the expansion appears reliable, radiative corrections can be quite large if $x \ll 1$. Although, it should be stressed that we have only calculated the rate to NLO for this model, and it is necessary to include two-loop contributions to ensure that the expansion converges. 

Moreover, perturbation theory might still work for smaller $x$ if vector-bosons are integrated out. This would give a cubic term in the leading-order potential, akin to equation \ref{eq:RadPotential}. Higher-order corrections should then be more well-behaved even if $x$ is small. They can still be large, though, as indicated by the radiative-barrier case in figure \ref{fig:yN}.

\sectioninline{Conclusion}
We find that radiative corrections to the nucleation rate can be large. By using a strict perturbative expansion, this paper calculates the size of these corrections for a variety of models. The calculations indicate that higher-order corrections are important, and should be included when studying extensions of the Standard Model. Even when perturbation theory is reliable, observables can change by a factor of $2$ once NLO corrections are included.

Our results also indicate that the perturbative expansion changes for strong transitions. This is because couplings and masses scale differently at the nucleation-scale for such transitions. Interestingly, it is not possible to make higher-order corrections arbitrarily suppressed for the models considered. This does not necessarily mean that perturbation theory is inadequate, but it does encourage caution when estimating the size of higher-order corrections to the rate.

In addition, this work illustrates the synergy between classical nucleation theory and effective high-temperature field theories~\cite{Gould:2021ccf}. Indeed, all non-equilibrium effects can be calculated within the effective theory, while temperature dependence is captured by effective couplings. 

For the radiative-barrier model, it is found that perturbation theory breaks down when $x\approx\frac{\lambda}{g^2}\sim 0.02$. This should be contrasted with $x\approx 0.1$, which is the endpoint of the first-order transition~\cite{Gurtler:1997hr, Rummukainen:1998as}. Meaning that perturbation theory only works in a narrow parameter-range for the radiative-barrier model.

Moreover, the results (see equation \ref{eq:ThinWallBehav}) indicate that perturbation theory breaks down in the large-bubble limit. A possible solution is to resum these large $R$ corrections into an effective theory as suggested by \cite{Gould:2021ccf}. However, it should be stressed that these cases often correspond to weak transitions, which might not be interesting gravitational-wave wise.

For future work it would be interesting to confirm that the calculations converge for radiative-barriers. This would require three-loop calculations, and would be the final calculable contribution due to the Linde problem~\cite{Linde:1980ts}.

Furthermore, the methods of this paper can be applied to models with two-step transitions, like for example singlet/triplet extensions of the Standard Model~\cite{Niemi:2021qvp,Bell:2020gug,Friedrich:2022cak}. Similar to the models studied in this paper, radiative corrections are expected to be sizeable for such extensions.

\vspace{3mm}
The author would like to thank Philipp Schicho, Oliver Gould, and Tuomas V.~I.~Tenkanen for insightful discussions and for a critical read-through of the manuscript. During the completion of this paper the author was made aware of similar methods used in the forthcoming paper \cite{Gould:2022lat}. 
This work has been supported by the Deutsche Forschungsgemeinschaft under Germany's Excellence Strategy - EXC $2121$ Quantum Universe - $390833306$; and by the Swedish Research Council, project number VR:$2021$-$00363$.

	\bibliography{Bibliography}%
\end{document}